\documentclass[prd,preprintnumbers,floatfix,aps,nofootinbib,notitlepage,showpacs]{revtex4}
\usepackage{amsmath}
\usepackage{graphicx}
\usepackage{epstopdf}
\usepackage{latexsym}
\usepackage{amssymb}

\newcommand\som[1]{ {\textstyle\sum\limits_{#1}}}

\begin{document}

\title{Assisted coupled quintessence}
\author{Luca Amendola}
\affiliation{Institut f\"ur Theoretische Physic, Universit\"at Heidelberg, Philosophenweg 16, D-69120 Heidelberg, Germany}
\author{Tiago Barreiro}
\affiliation{Departamento de Matem\'atica,  ECEO, Universidade Lus\'ofona de Humanidades e 
Tecnologias, Campo Grande, 376,  1749-024 Lisboa, Portugal}
\author{Nelson J. Nunes}
\affiliation{Faculty of Sciences and Centre for Astronomy and Astrophysics, University of Lisbon, 1749-016 Lisbon, Portugal}

\begin{abstract}
We study models of quintessence consisting of a number of scalar fields coupled  to several dark matter components.  In the case of exponential potentials the scaling solutions can be described in terms of a single field. The corresponding  effective logarithmic slope and effective coupling can be written in a simple form in terms of the individual slopes and couplings of the original fields. We also investigate  solutions where the scalar potential is negligible, in particular those leading to transient matter dominated solutions. Finally, we compute the evolution equations for the linear perturbations which will allow these models to be tested against current and future observational data.
\end{abstract}

\maketitle

\section{Introduction}

In the past decade, our understanding of the evolution of the Universe, its components
and respective abundances has increased to an unprecedented level. Results from various
independent observations of different scales have provided us with evidence for dark matter.
Supernovae combined with other independent observations suggest that the universe is currently 
undergoing accelerated expansion, possibly due to a negative pressure component
usually dubbed “dark energy”. We have also learned from analysis of the cosmic microwave
background radiation that the universe is close to flat and that the large scale structure 
developed through gravitational instability from a spectrum of adiabatic, nearly Gaussian
and nearly scale invariant density perturbations. These conclusions are consistent with the
predictions of the simplest inflationary paradigm, a short period of accelerated expansion
in the early Universe, introduced to explain the flatness, homogeneity and isotropy of the Universe.

Scalar fields are the most popular building blocks to construct candidate models of 
early Universe inflation and of the present day cosmological acceleration. They are appealing 
because such fields are ubiquitous in theories of high energy physics beyond the standard model. 
Models are usually constructed using a single field, however, there is also the interesting possibility that 
a cosmological behaviour arises from the presence of multiple scalar fields.
The idea that our Universe contains a vast number of 
light scalar fields is also based on expectations from landscape models, see e.g. ~\cite{2010PhRvD..81l3530A}.
Assisted inflation is an example of a model where many fields can cooperate to sustain inflation even if none 
is able to fuel it if evolving in isolation.  Assisted inflation for exponential potentials was proposed in
\cite{Liddle:1998jc} and extended in
\cite{Malik:1998gy, Copeland:1999cs} and the inclusion of a background fluid was
evaluated in Ref.~\cite{Coley:1999mj,Hartong:2006rt}. It was studied for
quadratic and quartic potentials in the context of  higher dimensional reduction
in Ref.~\cite{Kanti:1999vt,Kanti:1999ie,Kaloper:1999gm}, for Bianchi models
\cite{Aguirregabiria:2000hx,Aguirregabiria:2001gm},  for an ensemble of tachyon
fields \cite{Mazumdar:2001mm,Piao:2002vf,Singh:2006yy,Panigrahi:2007sq},  in
braneworld models \cite{Piao:2001dd,Panotopoulos:2007pg}, in particular string
theory realisations \cite{Lalak:2005hr,Ward:2005ti,Olsson:2007he} and taking
into account Loop Quantum Cosmology corrections \cite{Ranken:2012hp}.  A
multi-field dynamics was also studied in the context of Logamediate inflation
\cite{Barrow:2007zr} and k-inflation 
\cite{Ohashi:2011na}.

A set of several fields working together could also give rise to the accelerated
evolution of the Universe we currently observe. This is usually known as
assisted dark energy or assisted quintessence
\cite{Kim:2005ne,Tsujikawa:2006mw,Ohashi:2009xw,Karwan:2010xw}.

These fields could, of course, interact with the rest of the world and new  
forces between matter particles would arise. For common particles, these forces are tightly 
constrained by solar  system and gravitational experiments on Earth. Limits on these forces 
are not so strong for interactions involving neutrinos or dark matter and current bounds come from 
cosmological observations. 
Coupled quintessence, where a scalar field interacts with dark matter was
introduced in Refs.~\cite{Amendola:1999dr,Holden:1999hm,Amendola:1999er} and its
dynamics, properties and possible couplings was studied in
Refs.~\cite{Koivisto:2005nr,Gonzalez:2006cj,Valiviita:2008iv,Lee:2009ji,
Boehmer:2009tk,Majerotto:2009np,Valiviita:2009nu,LopezHonorez:2010ij,
  Tzanni:2014eja}. 

There  might be, however, more than one dark matter species, a suggestion made in Refs.~\cite{Khlopov:1995pa,Farrar:2003uw,Copeland:2003cv}.
The possibility that a scalar field is indeed coupled to more than a single
dark matter component has been raised recently in Ref.~\cite{Brookfield:2007au,Baldi:2012kt} and
the phenomenology of such a set up was investigated and compared to observations
in Refs.~\cite{Piloyan:2013mla,Piloyan:2014gta,Baldi:2014tja}. 

A natural extension of these works is, therefore, to investigate the
cosmological dynamics considering a group of several scalar fields coupled to
various dark matter components. Such a system has been addressed in
Ref.~\cite{Tsujikawa:2006mw}. Here  we extend that work with an explicit analysis of the background and of linear perturbations. In section II we introduce the general equations for $n$ scalar fields coupled to $m$ matter components. We then apply these to scalar potentials consisting of a sum of exponential terms in section III,  and an exponential of a sum of terms in section IV. We follow this in section V with the solutions where the scalar potential is negligible. In section VI we extend our analysis to the linear perturbations. Finally we conclude in section VII.

\section{General equations}

We consider an ensemble of $n$ scalar fields $\phi_i$ cross-coupled to an ensemble of $m$
dark matter components $\rho_\alpha$. The cross-couplings are described by the matrix $C_{i\alpha}$
where latin indexes $i,j$ identify  scalar field indexes and greek indexes $\alpha,\beta$ identify
the dark matter components. 
The equation of motion for the fields and the various dark matter components in a spatially flat 
Friedmann-Robertson-Walker metric with scale factor $a(t)$
is
then written as
\begin{eqnarray}
\ddot{\phi}_i + 3 H \dot{\phi}_i + V_{,\phi_i} &=& \kappa \sum_\alpha  C_{i\alpha} \rho_\alpha,
\\
\dot{\rho}_\alpha +3 H \rho_\alpha &=& - \kappa \sum_i C_{i\alpha} \dot{\phi}_i \rho_\alpha.
\end{eqnarray}
The solution for the dark matter component evolution can be given immediately in
terms of the values of the fields as
\begin{equation}
\rho_\alpha = {\rho_\alpha}_0  \exp\left(-3N - \kappa \sum_i C_{i\alpha} (\phi_i -
{\phi_i}_0)\right).
\end{equation}
The rate of change of the Hubble function is
\begin{equation}
\dot{H} = - \frac{\kappa^2}{2} \left( \sum_\alpha \rho_\alpha + \sum_i \dot\phi_i^2
\right) ,
\end{equation}
subject to the Friedmann constraint
\begin{equation}
H^2 = \frac{\kappa^2}{3} \left( \sum_\alpha \rho_\alpha + \sum_i \rho_{\phi_i} \right).
\end{equation}
where $\rho_{\phi_i} =  \sum_i \phi_i^2/2 + V(\phi_1,...,\phi_n)$.
In the next two sections we consider two possible forms for $V(\phi_1,...,\phi_n)$, both
leading to scaling solutions: a sum of exponential terms and an exponential of a sum of terms.

The dark matter field can in principle include the baryon component. However in this case the coupling will have to satisfy the 
strong solar system constraints and be in practice negligible.

\section{Sum of exponential terms: $V(\phi_1,...,\phi_n) = M^4 \sum_i e^{-\kappa
\lambda_i \phi_i}$ }

With the aim of finding the critical points of the evolution, we will rewrite
the equations of motion as a system of first order differential  equations. To
do this we define the new variables
\begin{eqnarray}
x_i \equiv \frac{\kappa \dot\phi_i}{\sqrt{6} H}, \hspace{1cm} y_i^2 \equiv
\frac{\kappa^2 V_i}{3 H^2}, \hspace{1cm} z_\alpha^2\equiv \frac{\kappa^2 \rho_\alpha}{3
H^2}, 
\end{eqnarray}
where $V_i = M^4 e^{-\kappa \lambda_i \phi_i}$. The evolution is now described
by
\begin{eqnarray}
\label{eqx1}
x_i' &=& -\left(3 + \frac{H'}{H}\right) x_i + \sqrt{\frac{3}{2}} \left(
\lambda_i y_i^2 + \sum_\alpha C_{i\alpha} z_\alpha^2\right), \\
\label{eqy1}
y_i'  &=& - \sqrt{\frac{3}{2}} \left( \lambda_i x_i +
\sqrt{\frac{2}{3}}\frac{H'}{H} \right) y_i, \\
\label{eqz1}
z_\alpha' &=&  - \sqrt{\frac{3}{2}} \left( \sum_i C_{i\alpha} x_i +  \sqrt{\frac{3}{2}} + 
\sqrt{\frac{2}{3}} \frac{H'}{H}\right) z_\alpha, \\
\label{eqdHH}
\frac{H'}{H} &=& -\frac{3}{2} \left( 1 + \sum_i (x_i^2 - y_i^2 ) \right), 
\end{eqnarray} 
where a prime means differentiation with respect to $N = \ln a$ and  the
Friedmann equation now reads
\begin{equation}
\label{friedmann}
\sum_i (x_i^2+y_i^2)+\sum_\alpha z_\alpha^2 = 1.
\end{equation}
Equation (\ref{eqdHH}) also defines the effective equation of state parameter
$w_{\rm eff}$, such that
\begin{equation}
\frac{H'}{H} = -\frac{3}{2} (1+ w_{\rm eff}), 
\end{equation}
and then,
\begin{equation}
w_{\rm eff} = \sum_i (x_i^2 - y_i^2 ).
\end{equation}

\subsection{Scalar field dominated solution}
We will start by reviewing the case  in which we can neglect the matter contributions
($z_\alpha = 0$) studied in the context of Assisted Inflation. Using
Eqns.~(\ref{eqy1}), (\ref{eqdHH}) and (\ref{friedmann}), we obtain that 
\begin{equation}
x_i = \frac{1}{\sqrt{6}} \frac{1}{\lambda_i \sum_j 1/\lambda_j^2}\,.
\end{equation}
By using the Friedmann equation  the effective equation of
state is obtained as
\begin{equation}
w_{\rm eff} = -1 + \frac{1}{3} \lambda_{\rm eff}^2, 
\end{equation}
where the effective slope, $\lambda_{\rm eff}$, given by
\begin{equation}
\frac{1}{\lambda_{\rm eff}^2} = \sum_i \frac{1}{\lambda_i^2},
\end{equation} 
describes the corresponding
logarithmic slope in the potential needed to replicate the dynamics with only
one field.
This means that  increasing the number of fields makes 
$\lambda_{\rm eff}$ smaller and, therefore, easier to obtain an accelerated expansion even if the individual slopes are too large to fuel it if acting in isolation. This effective slope $\lambda_{\rm eff}$ is similar to the result obtained in the assisted inflation scenarios that were studied in \cite{Liddle:1998jc,Malik:1998gy,Copeland:1999cs}.

\subsection{Scaling solution}
The system of equations has many fixed points, but we are particularly
interested in the case when all the variables $x_i$, $y_i$ and $z_\alpha$ are
non-vanishing since these are the most interesting in cosmology. Using Eqns.~(\ref{eqy1}) and (\ref{eqz1}) and considering for a moment only two fields, the fixed points of
the system are at
\begin{eqnarray}
\label{x1a}
x_1 &=& \sqrt{\frac{3}{2}} \frac{1}{\lambda_1-\gamma_1 }, \\
\label{x2a}
x_2 &=&  \sqrt{\frac{3}{2}} \frac{1}{\lambda_2-\gamma_2 }.
\end{eqnarray}
where 
\begin{eqnarray}
\gamma_1 &=& C_{11}+C_{21}\frac{\lambda_{1}}{\lambda_{2}}, \\
\gamma_2 &=& C_{22}+C_{12}\frac{\lambda_{2}}{\lambda_{1}}.
\end{eqnarray}
Equations (\ref{eqz1}) also provide the constraint  
\begin{equation}
\frac{x_2}{x_1} = \frac{\lambda_1}{\lambda_2} =
\frac{C_{12}-C_{11}}{C_{21}-C_{22}}.
\end{equation}
and $\gamma_1/\lambda_1 = \gamma_2/\lambda_2$.

\begin{figure}[!ht]
\begin{center}
\includegraphics[width=0.7\columnwidth]{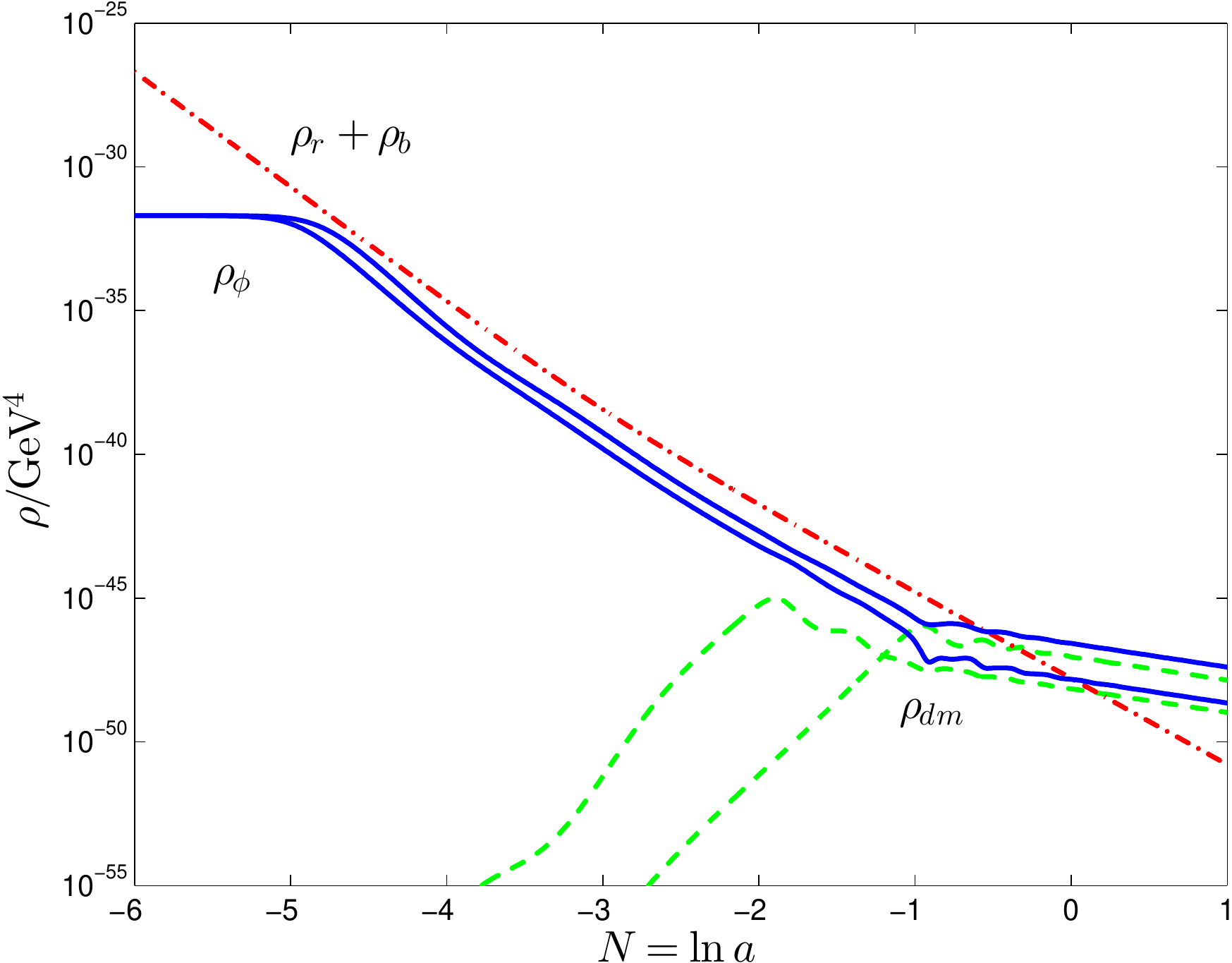}
\caption{\label{fig:pot1} Evolution of energy densities for the sum of
exponentials potential with $\lambda_1 = 10$, $\lambda_2 = 5.4$, $C_{11}= 90$,
$C_{12}= -8$, $C_{21}= -63$, $C_{22}= -10$.}
\end{center}
\end{figure}

We can now compute the effective equation of state parameter of the Universe
knowing that 
\begin{equation}
\frac{H'}{H} = -\sqrt{\frac{3}{2}} \lambda_i x_i =   -\frac{3}{2} (1+w_{\rm
eff})
\end{equation}
it is obtained that
\begin{equation}
w_{\rm eff} = -1 + \frac{1}{2}\sqrt{\frac{2}{3}} (\lambda_1 x_1  +  \lambda_2
x_2).
\end{equation}
Substituting for $x_1$ and $x_2$ as found earlier, we can write 
\begin{equation}
w_{\rm eff} = \frac{\gamma_i}{\lambda_i - \gamma_i},
\end{equation}
where $i = 1,2$. Making use of the property that  $\gamma_1/\lambda_1 =
\gamma_2/\lambda_2$ we can define and effective coupling of the system such that
\begin{equation}
C_{\rm eff} \equiv  \lambda_{\rm eff} \frac{\gamma_i}{\lambda_i},
\end{equation}
to write
\begin{equation}
\label{eqweff}
w_{\rm eff} =  \frac{C_{\rm eff}}{\lambda_{\rm eff}-C_{\rm eff}},
\end{equation}
Moreover, $\Omega_\phi = x_1^2+x_2^2 + y_1^2+ y_2^2$ and given the equality
$w_{\rm eff}= x_1^2+x_2^2-y_1^2- y_2^2$, it results that the total contribution
of the fields to the total energy budget is
\begin{equation}
\Omega_{\phi} = 1 + 2(x_1^2+ x_2^2)-  \frac{1}{2} \sqrt{\frac{2}{3}} (\lambda_1
x_1 + \lambda_2 x_2).
\end{equation}
Substituting now for $x_1$ and $x_2$ we obtain that 
\begin{equation}
\label{eqOmegaphieff}
\Omega_\phi = \frac{3-\lambda_{\rm eff} C_{\rm eff}+C_{\rm eff}^2}{(\lambda_{\rm
eff}-C_{\rm eff})^2},
\end{equation}
where the effective $\lambda_{\rm eff}$ is given by
\begin{equation}
\frac{1}{\lambda_{\rm eff}^2} = \frac{1}{\lambda_1^2}+\frac{1}{\lambda_2^2}.
\end{equation}
A typical evolution of the energy densities is illustrated in
Fig.~\ref{fig:pot1}.    

\begin{figure}[!ht]
\begin{center}
\includegraphics[width=0.7\columnwidth]{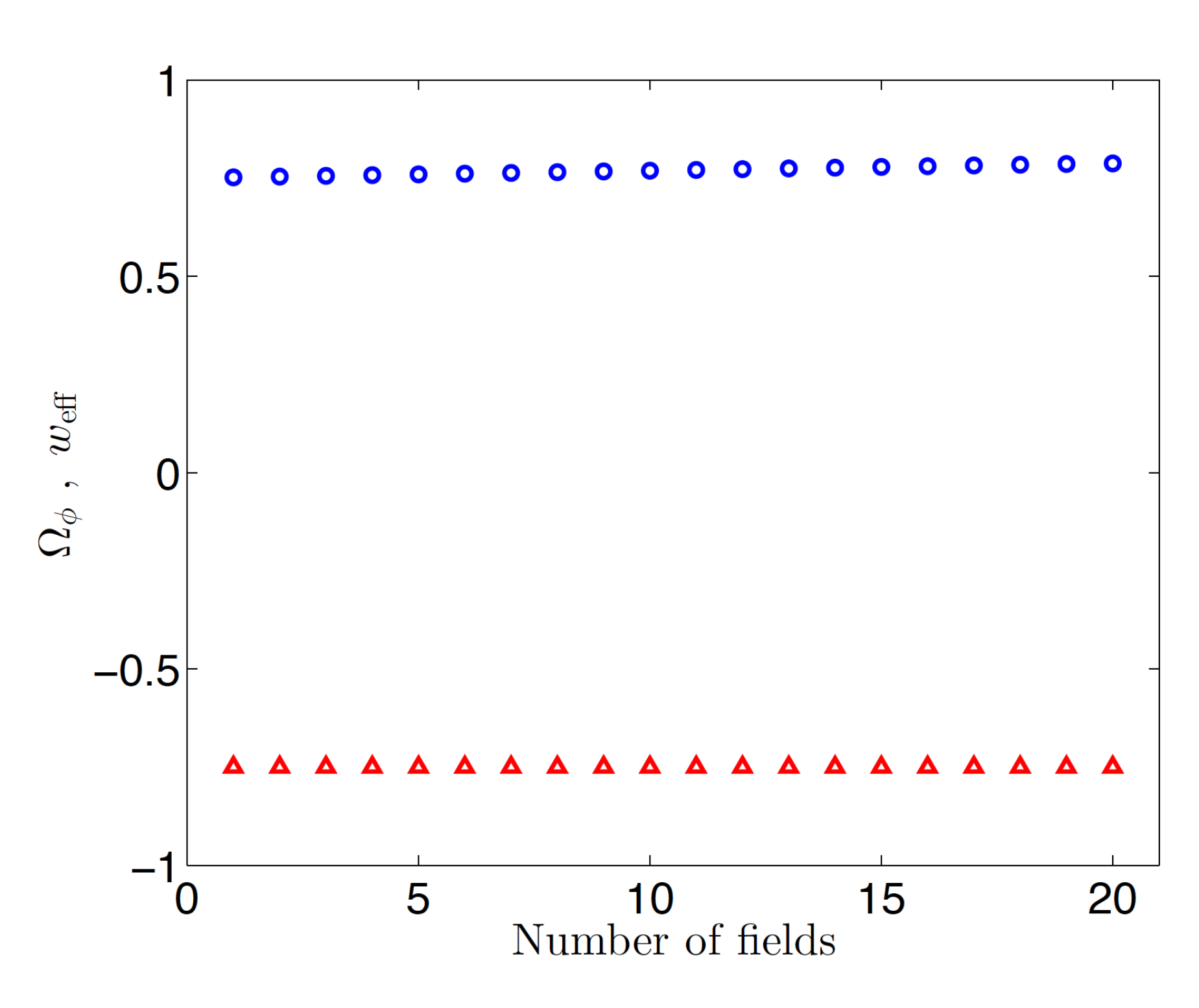}
\caption{\label{fig:eff1} Dependence of $w_{\rm eff}$ (triangles) and
$\Omega_\phi$ (circles) with the number of fields in the case when all fields
are replicas of the same field and for the sum of exponentials potential. We
used $\lambda_1 = 10$ and  $C_{11}=-30$.}
\end{center}
\end{figure}

The generalization to more than two fields is fairly trivial for this potential.
Eqns. (\ref{x1a}) and (\ref{x2a}) are still valid for $n$ fields $\times$  $m$
dark matter components and in general we have,
\begin{equation}
\frac{\gamma_i}{\lambda_i}  = \sum_j^n \frac{C_{j1}}{\lambda_j} = ... = \sum_j^n
\frac{C_{jm}}{\lambda_j}
\end{equation}
for every $i= 1,...,n$.
It is also instructive to consider how these quantities simplify when we
consider that all the fields are just replicas of, say, the field $\phi_1$. 
By this we mean that the matrix $C$ is diagonal with entries $C_{ii} = C_{11}$
and $\lambda_i = \lambda_1$ for all indexes $i$. It is easy to verify that in
this case $\lambda_{\rm eff} = \lambda_1/\sqrt{n}$ and $C_{\rm eff} =
C_{11}/\sqrt{n}$ and consequently 
\begin{equation}
w_{\rm eff} =  \frac{C_{11}}{\lambda_1-C_{11}},
\end{equation}
takes the same value as for a single field system, however,
\begin{equation}
\Omega_\phi =  \frac{3 n -\lambda_1 C_{11}+C_{11}^2}{(\lambda_1-C_{11})^2},
\end{equation}
has a slight dependence on the number of fields $n$ as illustrated in
Fig.~\ref{fig:eff1}.

\section{Exponential of a sum of terms: $V(\phi_1,...,\phi_n) = M^4 e^{-\sum_i
\kappa \lambda_i \phi_i}$ }

For this case we only need to define one single $y$ such that, 
\begin{eqnarray}
x_i \equiv \frac{\kappa \dot\phi_i}{\sqrt{6} H}, \hspace{1cm} y^2 \equiv
\frac{\kappa^2 V}{3 H^2}, \hspace{1cm} z_\alpha^2\equiv \frac{\kappa^2 \rho_\alpha}{3
H^2}.
\end{eqnarray}
The evolution is now described by
\begin{eqnarray}
\label{eqx2}
x_i' &=& -\left(3 + \frac{H'}{H}\right) x_i + \sqrt{\frac{3}{2}} \left(
\lambda_i y^2 + \sum_\alpha C_{i\alpha} z_\alpha^2\right), \\
\label{eqy2}
y'  &=& - \sqrt{\frac{3}{2}} \left( \sum_i \lambda_i x_i +
\sqrt{\frac{2}{3}}\frac{H'}{H} \right) y, \\
\label{eqz2}
z_\alpha' &=&  - \sqrt{\frac{3}{2}} \left( \sum_i C_{i\alpha} x_i +  \sqrt{\frac{3}{2}} + 
\sqrt{\frac{2}{3}} \frac{H'}{H}\right) z_\alpha, \\
\frac{H'}{H} &=& -\frac{3}{2} \left( 1 + \sum_i x_i^2  - y^2 \right), 
\end{eqnarray} 
where the Friedmann equation now reads
\begin{equation}
\label{friedmann2}
\sum_i x_i^2+y^2+\sum_\alpha z_\alpha^2 = 1.
\end{equation}

\subsection{Scalar field dominated solution}
When the matter components are negligible ($z_\alpha = 0$), we can use
Eqns.~(\ref{eqy2}), (\ref{eqdHH}) and (\ref{friedmann2}) to obtain
\begin{equation}
x_i = \frac{\lambda_i}{\sqrt{6}},
\end{equation}
and then the equation of state parameter reads
\begin{equation}
w_{\rm eff} = -1 + \frac{1}{3} \lambda_{\rm eff},
\end{equation}
where the effective slope, $\lambda_{\rm eff}$ is now
\begin{equation}
\lambda_{\rm eff}^2 = \sum_i \lambda_i^2.
\end{equation}
For this case, increasing the number of fields increases the value of
$\lambda_{\rm eff}$ and an accelerated expansion becomes more difficult to be attained. Again, this mimics previous results obtained in an assisted inflation setting in \cite{Malik:1998gy,Copeland:1999cs}.

\subsection{Scaling solution}
In order to find the critical points corresponding to the scaling solution, when
all the variables $x_i$, $y$ and $z_\alpha$ are non-vanishing, we could proceed as
for the previous potential. However, it is considerably easier to first start
with a redefinition of variables. 
Let us observe that the evolution equations are invariant under an orthogonal
transformation of $x_i$, $\lambda_i$ and $C_{ij}$. That is,
\begin{eqnarray}
\hat{x}_i &=& Q_{ij} x_j, \label{ortrans1} \\
\hat{\lambda}_i &=& Q_{ij} \lambda_j, \label{ortrans2}  \\
\hat{C}_{ij} &=& Q_{il} C_{lj}, \label{ortrans3} 
\end{eqnarray}
were $Q_{ij}$ is an orthogonal matrix, {\it i.e.}, $Q_{il} Q^T_{lj} =
Q_{il}Q_{jl} = \delta_{ij}$.

We will now give a working example for two fields and two dark matter
components. We will show that, in the case of this potential it is always
possible to rotate the fields such that $\hat x_2 = 0$. Looking for the case
where all the variables $\hat x_i$, $\hat y$ and $\hat z_i$ are non-vanishing,
from Eqs.~(\ref{eqy2}) and (\ref{eqz2}), we obtain the conditions
\begin{eqnarray}
(\hat \lambda_1 - \hat C_{11} ) \hat x_1 + (\hat \lambda_2 - \hat C_{21}) \hat
x_2 &=& \sqrt{\frac{3}{2}}, \\
(\hat \lambda_1 - \hat C_{12} ) \hat x_1 + (\hat \lambda_2 - \hat C_{22}) \hat
x_2 &=& \sqrt{\frac{3}{2}}.
\end{eqnarray}
It is easy to show that when $\hat C_{11} = \hat C_{12}$ the solution yields
$\hat x_2 = 0$. Using the orthogonal transformation 
(\ref{ortrans2}) with the condition  $\hat C_{11} = \hat C_{12}$,
\begin{equation}
 Q_{11} C_{11} + Q_{12} C_{21} =  Q_{11} C_{12} + Q_{12} C_{22}
\end{equation}
together with the constraint that the rows of $Q$ are unit vectors, $Q_{11}^2 +
Q_{12}^2 = 1$,
 gives us the result
\begin{align}
Q_{11} & = \frac{C_{22} - C_{21}}{\sqrt{(C_{11}-C_{12})^2 + (C_{22} -
C_{21})^2}}, \\
Q_{12} & = \frac{C_{11} - C_{12}}{\sqrt{(C_{11}-C_{12})^2 + (C_{22} -
C_{21})^2}}
\end{align}
where we have chosen the positive roots. 
Since $\hat{x}_2 = 0$ we do not need to compute the rest of the matrix $Q$. We
can now obtain the effective coupling
\begin{eqnarray}
C_{\rm eff} = \hat C_{11} &=& Q_{11} C_{11} + Q_{12} C_{21}  \\
& =& \frac{C_{22} C_{11}-C_{21} C_{12}}{\sqrt{(C_{11}-C_{12})^2 + (C_{22} -
C_{21})^2}}
\end{eqnarray}
and the effective $\lambda_{\rm eff}$,
\begin{eqnarray}
\lambda_{\rm eff}  = \hat{\lambda}_{1} &=& Q_{11} \lambda_1 + Q_{12} \lambda_2
\\
& =&  \frac{(C_{22}-C_{21}) \lambda_1 + (C_{11}-C_{12})
\lambda_2}{\sqrt{(C_{11}-C_{12})^2 + (C_{22} - C_{21})^2}}
\end{eqnarray}
The scaling solution can then be written as
\begin{eqnarray}
\label{hatx1}
\hat{x}_1 &=& \sqrt{\frac{3}{2}} \frac{1}{\lambda_{\rm eff} - C_{\rm eff}}, \\
\hat{x}_2 &=& 0.
\end{eqnarray}
We can then compute the effective equation of state parameter and the total
scalar field contribution from the expressions
\begin{eqnarray}
\label{weff2}
w_{\rm eff} &=& -1 + \sqrt{\frac{2}{3}} (\hat \lambda_1\hat x_1+ \hat \lambda_2
\hat x_2) = \frac{C_{\rm eff}}{\lambda_{\rm eff}-C_{\rm eff}}, \\
\label{Omegaeff2}
\Omega_\phi &=& 1 + 2(\hat x_1^2 + \hat x_2^2) - \sqrt{\frac{2}{3}} (\hat
\lambda_1\hat x_1+ \hat \lambda_2 \hat x_2) = \frac{3-\lambda_{\rm eff} C_{\rm
eff}+C_{\rm eff}^2}{(\lambda_{\rm eff}-C_{\rm eff})^2}.
\end{eqnarray}

It is straightforward  to extend these results to the case of $n$ scalar fields
and $n$ dark matter components. We can  have 
$\hat x_i = 0$ for $i \geq 2$ provided that
\begin{align}
\hat C_{ii} = \hat C_{ij} \text{ for } j \geq i.
\end{align}
This can always be achieved through the orthogonal transformation
Eqs~(\ref{ortrans1})--(\ref{ortrans3}).
Again we only really need to obtain the first row $Q_{1j}$, of the $Q$ matrix to
obtain the value of the effective coupling $C_{\rm eff}$. This comes through the
solution to the equation
\begin{equation}
C_{\rm eff} = \hat C_{1i} = \sum_j Q_{1j} C_{j1},
\end{equation}
for any $i$,  together with the unitarity condition, $Q_{1i}^2 = 1$. Once we
have $Q_{1i}$ we can obtain the effective slope $\lambda_{\rm eff}$
\begin{equation}
\lambda_{\rm eff} = \hat \lambda_1 = \sum_j Q_{1j} \lambda_j.
\end{equation}
and $\hat x_1$, $\Omega_\phi$ and $w_{\rm eff}$ will be given by the same 
Eqs.~(\ref{hatx1})--(\ref{Omegaeff2}).

To see how this would work, let us look at the simple case of a diagonal
coupling matrix $C$. In this case, we get for $Q_{1i}$
\begin{equation}
Q_{1i} = \frac{1}{C_{ii} \sqrt{\sum_l 1/C_{ll}^2}}, 
\end{equation}
such that
\begin{eqnarray}
\frac{1}{C_{\rm eff}^2} & =&  \sum_i \frac{1}{C_{ii}^2}, \\
\lambda_{\rm eff} & =& C_{\rm eff} \sum_i \frac{\lambda_i}{C_{ii}}.
\end{eqnarray}

\begin{figure}[h!]
\begin{center}
\includegraphics[width=0.7\columnwidth]{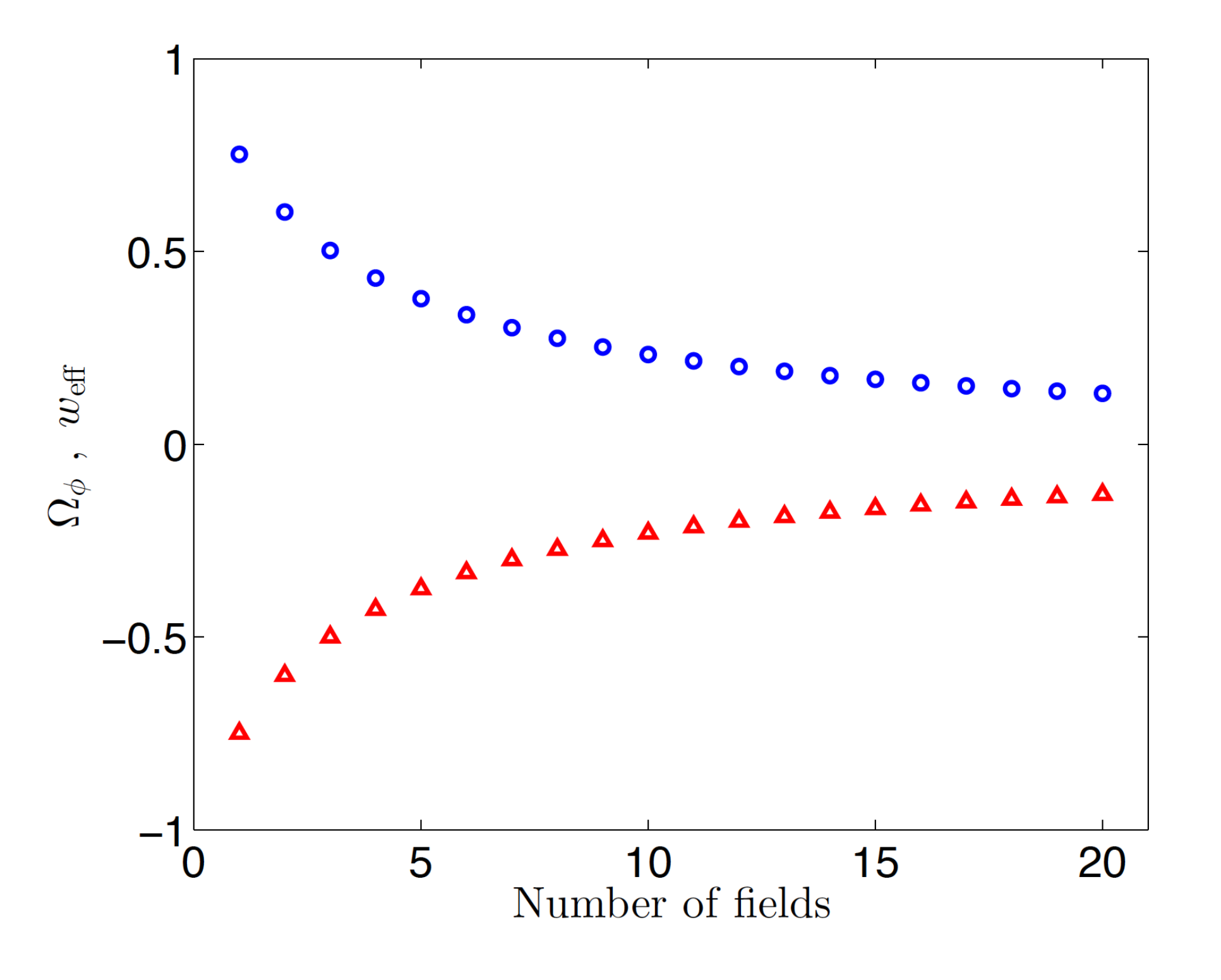}
\caption{\label{fig:eff2} Dependence of $w_{\rm eff}$ (triangles) and
$\Omega_\phi$ (circles) with the number of fields in the case when all fields
are replicas of the same field and for the exponential of sum of fields
potential. We used $\lambda_1 = 10$ and  $C_{11}=-30$.}
\end{center}
\end{figure}

When all the fields are a copy of field $\phi_1$, then the expressions become
fairly simple and give $C_{\rm eff} = C_{11}/\sqrt{n}$ and $\lambda_{\rm eff} =
\sqrt{n}\lambda_1$. For this potential both $w_{\rm eff}$ and $\Omega_\phi$ have
a strong dependence on the number of fields
\begin{eqnarray}
w_{\rm eff} &=& \frac{C_{11}}{n \lambda_1 - C_{11}}, \\
\Omega_\phi &=& \frac{3n - C_{11} \lambda_1 n +
C_{11}^2}{(n\lambda_1-C_{11})^2},
\end{eqnarray}
which is illustrated in 
Fig.~\ref{fig:eff2} for $\lambda_1 = 10$ and  $C_{11}=-30$.

\section{Scalar potential independent solutions}

We can find fixed point solutions where the scalar potential energy density is negligible, so these solutions are independent of the type of scalar potential used. These can be divided into three different types.

\subsection{Subdominant potential solution}
This critical point corresponds to the case where the scalar potential is
negligible and the total energy density is comprised of the scalar fields kinetic energies and the dark matter components. We can use either Eq.~(\ref{eqz1}) or (\ref{eqz2}) and the Friedmann
constraint to obtain
\begin{equation}
x_i = \sqrt{\frac{2}{3}} C_{i\alpha},
\end{equation}
for any $\alpha$. This immediately gives the simple expression for the equation of
state parameter and the total scalar field contribution
\begin{equation}
w_{\rm eff} = \Omega_\phi = \sum_i x_i^2 = \frac{2}{3} \sum_i C_{i\alpha}^2.
\end{equation}

\subsection{Kinetic dominated solution}
This critical point occurs when only the kinetic energy of the fields is
non-vanishing. It is immediately obtained that $\sum_i x_i^2 = 1$ and
\begin{equation} 
w_{\rm eff} = \Omega_\phi = 1.
\end{equation}

\subsection{Matter dominated solution}
The matter dominated solution is the fixed point characterized by $x_i $,
$y_i=0$, (or $y=0$) and
\begin{equation}
\sum_\alpha C_{i\alpha} z_\alpha^2 = 0,
\end{equation}
which essentially means that the partial derivative with respect to the field
$\phi_i$ of the sum of all 
the dark matter contributions must vanish. This solution was discussed in a number of recent publications 
for a single field system and two dark matter components with symmetric couplings  \cite{Baldi:2012kt,Piloyan:2013mla,Piloyan:2014gta,Baldi:2014tja}. 
For the simple system of two fields
and two dark matter components we obtain that the field must settle at the
bottom of a valley defined by the flat direction in the $\phi_1$--$\phi_2$
plane,
\begin{eqnarray}
(C_{11}-C_{12})(\phi_1-{\phi_1}_0) + (C_{21}-C_{22})(\phi_2-{\phi_2}_0) &=&
\frac{1}{\kappa} \ln \left(- \frac{C_{11}}{C_{12}} 
\frac{{\rho_1}_0}{{\rho_2}_0} \right)  \nonumber \\
&=& \frac{1}{\kappa} \ln \left(- \frac{C_{21}}{C_{22}} 
\frac{{\rho_1}_0}{{\rho_2}_0} \right).
\end{eqnarray}
For consistency, we observe that the couplings must satisfy the simple relation
\begin{equation}
\frac{C_{11}}{C_{12}} = \frac{C_{21}}{C_{22}},
\end{equation}
otherwise the flat direction is non-existent and the field will keep on
evolving.

\begin{figure}[t]
\begin{center}
\includegraphics[width=0.7\columnwidth]{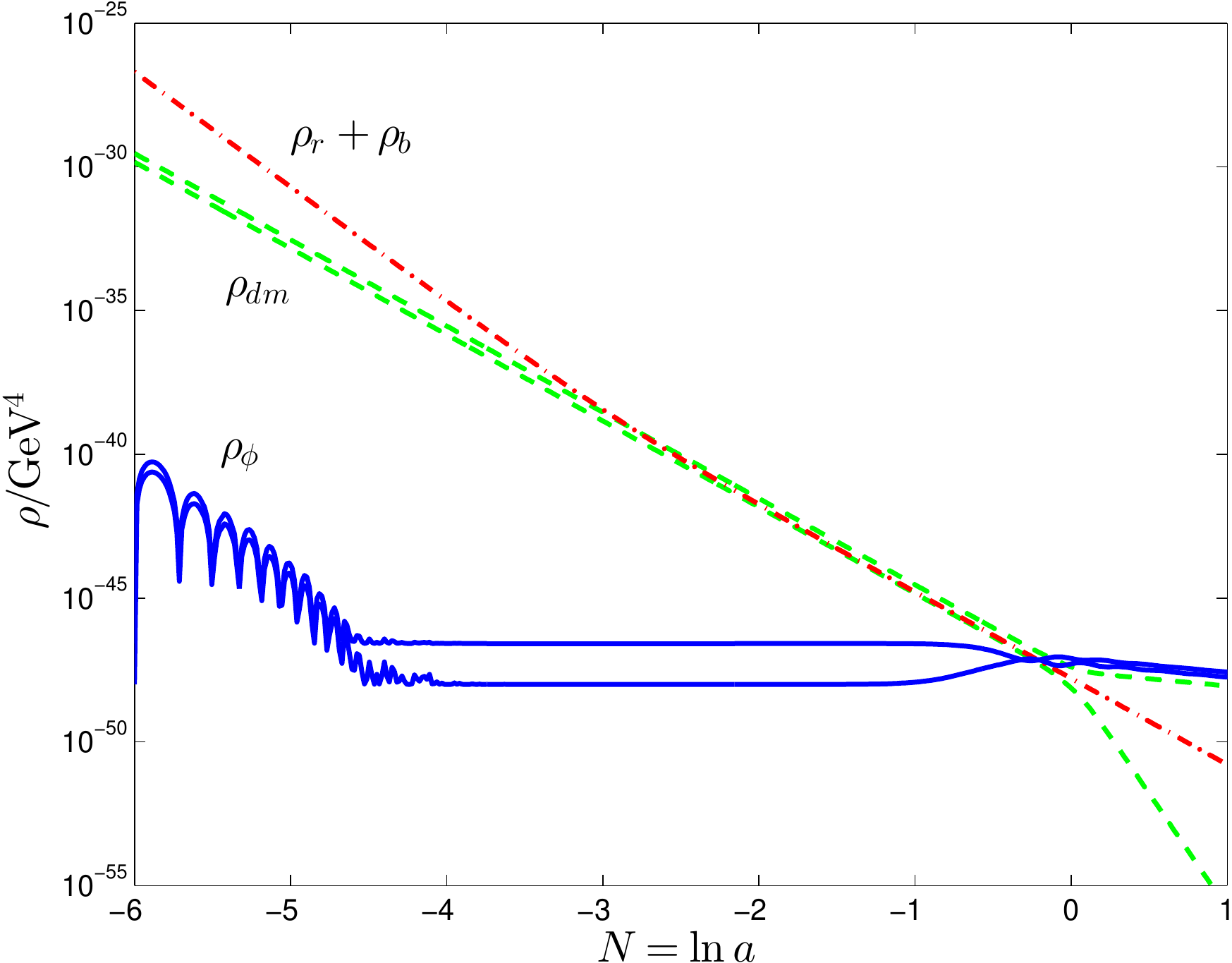}
\caption{\label{fig:pot1dmdom} Example of a matter dominated solution followed
by a scaling solution by only one of the dark matter components. Evolution of
energy densities for the sum of exponentials potential with $\lambda_1 = 10$,
$\lambda_2 = 10$, $C_{11}= -20$, $C_{12}= 40$, $C_{21}= -30$, $C_{22}= 60$. The
initial condition for the scalar fields is very near the bottom of the effective
potential which causes the initial oscillations before $N = -4$.}
\end{center}
\end{figure}

This solution is necessarily unstable. As the dark matter contribution decays
with $e^{-3N}$, eventually the scaling solution or the scalar field dominated
solution becomes more important.  Such an example is illustrated in
Fig.~\ref{fig:pot1dmdom}. This critical point allows the dark matter to have a
substantial contribution before the scaling regime and before the Universe
accelerates. 

\section{Linear density perturbations}

We now extend our analysis to linear perturbations. This would allow to test the multi-coupled model with current
and future data on galaxy clustering.
We perturb  the flat-space metric as
\begin{equation}
ds^2 = -(1+2\Psi){\rm d}t^2 + a^2(1-2\Phi) \eta_{ij}{\rm d} x^i {\rm d} x^j
\end{equation}
where $\Phi,\Psi$  are functions of space and time, and expand the fields as $\phi_i \rightarrow \phi_i+ \delta \phi_i$.

The energy-momentum tensor for the various fields can be written as
\begin{equation}
 T_{\mu\nu}^{\phi} = \sum_i \partial_\mu \phi_i \partial_\nu \phi_i + 
 g_{\mu\nu}\left(-\frac{1}{2} g^{\rho\sigma} \sum_i \partial_\rho \phi_i
\partial_\sigma \phi_i - V(\phi_1,...,\phi_n)\right),
 \end{equation}
 and for the component $\alpha$ of dark matter as
 \begin{equation}
 T_{\mu\nu}^{\mathrm{dm} \,(\alpha)} = \sum_\alpha \rho_\alpha u_\mu^{(\alpha)} u_\nu^{(\alpha)},
\end{equation}
where $u^{(\alpha)}_{\mu}$ is the four-velocity of component $\alpha$ of the matter fluid.
The conservation equations then read
\begin{eqnarray}
\nabla_\mu T^{\phi}~^\mu\!_\nu &=& -\kappa \sum_{i}
 \left( \som{\alpha} C_{i\alpha} \rho_\alpha \right) \nabla_\nu \phi_i  \\
  \nabla_\mu T^{\mathrm{dm} (\alpha)}~^\mu\!_\nu &=&
      \kappa \left( \som{i} C_{i\alpha} \nabla_\nu \phi_i \right) \rho_\alpha.
 \end{eqnarray}
We want to obtain the equations for the linear perturbations around the background values for the fields, $\bar{\phi}_i$ and for the matter components $\bar{\rho}_\alpha$. We define the field perturbations as $\delta \phi_i = \phi_i - \bar{\phi}_i$ and the matter density contrasts as 
$\delta_\alpha = (\rho_\alpha-\bar{\rho}_\alpha)/\bar{\rho}_\alpha$.

Going to Fourier space, we obtain the following equations of motion for the $\phi_i$  field perturbations of wavenumber $k$
\begin{equation}
\delta \ddot{\phi}_i + 3 H \delta \dot{\phi}_i
  + \som{j} V_{,\phi_i \phi_j} \delta\phi_j
  + \frac{k^2}{a^2} \delta \phi_i 
+ 2 V_{,\phi_i} \Phi -4 \dot{\phi}_i \dot{\Phi}
-2 \kappa \som{\alpha} C_{i\alpha} \rho_\alpha  \Phi
- \kappa \som{\alpha} C_{i\alpha} \rho_\alpha \delta_\alpha = 0,
\end{equation}
and for the density contrast of the matter component  $\alpha$,
\begin{equation}
\dot{\delta}_{\alpha} + \frac{\theta_\alpha}{a} - 3 \dot{\Phi}
+ \kappa \som{i}  C_{i\alpha} \delta \dot{\phi}_i = 0, 
\end{equation}
where $\theta_\alpha = \vec{\nabla} \cdot \vec{v}_\alpha$ and $\vec{v}_\alpha$ is the velocity of the $\alpha$ component of the matter fluid.

By differentiation we then get,
\begin{align}
\label{phiddot}
\ddot{\delta}_\alpha
& + 3 \ddot{\Phi}
+ \kappa \som{i} C_{i\alpha} \delta \ddot{\phi}_{i}
+ \frac{k^2}{a^2} \left(\Phi -\som{i} C_{i\alpha}\delta \phi_{i} \right)
+ (2 H - \kappa \som{i} C_{i\alpha} \dot{\phi}_{i})
   \left(\dot{\delta}_\alpha - 3 \dot{\Phi} - \kappa \som{i} C_{i\alpha} \delta \dot{\phi}_{i} \right) = 0.
\end{align}

The $ij$ component of Einstein's equations for $i\neq j$ yields
\begin{equation}
\Psi = \Phi,
\end{equation}
and using this equality from now on, the $00$ component gives
\begin{align}
\label{deltaddot}
&3H (\dot\Phi+H\Phi)
+ \tfrac{1}{2} \kappa^2 \som{i} \left(
 \dot\phi_i \delta\dot\phi_i
+ V_{,\phi_i} \delta\phi_i
-\dot\phi_i^2 \Phi  \right)
+\tfrac{1}{2} \kappa^2 \som{\alpha} \rho_\alpha \delta_\alpha
+ \frac{k^2}{a^2} \Phi = 0.
\end{align}
Combining Eqs.~(\ref{phiddot}) and (\ref{deltaddot}) and working  in the limit of small scales, {\it s.t.}, $(k/a)^2 \gg H^2$,
 it turns out  that the equation of motion for the linear matter perturbations is

\begin{align}
\ddot\delta_\alpha + 2H \left(1-\tfrac{1}{2} \kappa \som{i} C_{i\alpha} \frac{\dot\phi_i}{H} \right) \dot\delta_\alpha
- \tfrac{1}{2} \kappa^2 \sum_{\beta} (1 + \som{i} C_{i\alpha} C_{i\beta}) \rho_\beta \delta_\beta = 0.
\end{align}

Using as time variable, $N = \ln a$, we can rewrite  these equations of motion in  the form
\begin{eqnarray}
\delta_\alpha''
+ \Big( 2 -\tfrac{3}{2} \som{\beta} \Omega_\beta
  - \som{i} ( \tfrac{1}{2} \kappa^2\phi_i'^2 + C_{i\alpha} \kappa \phi_i')  \Big)   \delta_\alpha'
- \tfrac{3}{2} \sum_\beta ( 1 + \som{i} C_{i\alpha} C_{i\beta} ) \Omega_\beta \delta_\beta = 0,
\end{eqnarray}
which can also be written, using the definition of $x_i$, as
\begin{eqnarray} \label{multilin}
\delta_\alpha'' +\Big(2 -\tfrac{3}{2} \som{\beta} \Omega_\beta
  - \som{i} ( 3 x_i^2 + \sqrt{6}C_{i\alpha}  x_i )  \Big) \delta_\alpha'
- \tfrac{3}{2} \sum_\beta ( 1 + \som{i} C_{i\alpha} C_{i\beta} ) \Omega_\beta \delta_\beta = 0.
\end{eqnarray}

It should be clear that for a given $\alpha$ component a too large positive or negative $\sum_i C_{i\alpha} C_{i\beta}$, might imply, respectively,  a  very strong growth or damping of $\delta_\alpha$, a situation that must be avoided.

For the case of two scalar fields coupling with two components of dark matter, the equations for the matter linear perturbations become
\begin{align}
\delta_1'' &+\big(2 -\tfrac{3}{2} (\Omega_1 + \Omega_2)
  -  3 (x_1^2 + x_2^2)
 - \sqrt{6} (C_{11}  x_1 + C_{21} x_2)  \big) \delta_1' \nonumber\\ \qquad
&- \tfrac{3}{2}  ( 1 + C_{11}^2 + C_{12}^2 ) \Omega_1 \delta_1
- \tfrac{3}{2}  (1 + C_{11} C_{12} + C_{21} C_{22}) \Omega_2 \delta_2 = 0.
\intertext{ and }
\delta_2'' &+\big(2 -\tfrac{3}{2} (\Omega_1 + \Omega_2)
  -  3 (x_1^2 + x_2^2)
 - \sqrt{6} (C_{12}  x_1 + C_{22} x_2)  \big) \delta_1' \nonumber\\ \qquad
&- \tfrac{3}{2}  ( 1 + C_{12} C_{11} + C_{22} C_{21} ) \Omega_1 \delta_1
- \tfrac{3}{2}  (1 + C_{12}^2 + C_{22}^2) \Omega_2 \delta_2 = 0.
\end{align}

That is, for a scaling solution they are of the form
\begin{eqnarray}
\delta_1'' + A_1 \delta_1' - B_1 \delta_1 - B_2 \delta_2 &=& 0, \\
\delta_2'' + A_2 \delta_2' - C_1 \delta_1 - C_2 \delta_2 &=& 0,
\end{eqnarray}
where $A_j$, $B_j$ and $C_j$ are all constants.

The solutions for these coupled equations can be written as $\delta_1 \propto e^{\xi N}\propto a^\xi$ and $\delta_2 = b \delta_1$, however, the analytical relations between $\xi$ and $b$ in terms of the $A_{j}$, $B_{j}, C_{j}$ are too complicated to be of any practical  use. 
In the cases when $b \ll B_1/B_2$ and $b \ll C_1/C_2$ we can substitute the solutions in the above equations and obtain a set of equations to estimate $\xi$ and $b$ given by
\begin{eqnarray}
\xi^2 + A_1 \xi - B_1 &\approx& 0, \\
\xi^2 + A_2 \xi - C_1/b &\approx& 0,
\end{eqnarray}
which give the analytical results
\begin{eqnarray}
\xi &\approx& -\frac{1}{2}  \left(A_1 \pm \sqrt{A_1^2+ 4B_1} \right), \\
b &\approx& \frac{1}{2} \frac{C_1}{B_1^2 + B_1 A_2 (A_1-A_2)} \left[ 2 B_1 + (A_1-A_2) \left(A_1 \pm \sqrt{A_1^2+4B_1}\right) \right].
\end{eqnarray}
Of course this approximation is only reliable  when the assumptions $b \ll B_1/B_2, C_1/C_2$ are indeed verified.  This happens for instance when $\Omega_2\ll\Omega_1$, i.e. when one of the dark matter component is much smaller than the other.

The  growth rate $f\equiv d\log\delta/d\log a = \xi$ can be then directly compared to observations. This will be performed in future work. 


\section{Conclusions}
We have studied a cosmological system composed of a set of scalar fields coupled to an ensemble of dark matter components, thus generalizing previous work. The obtained solution can either be applied to construct early Universe inflationary solutions or provide a mechanism to explain the current accelerated expansion of the Universe and the ratio of abundances between dark energy and dark matter. More specifically, we have investigated two representative types of potential leading to analytical solutions. 
We have seen that the scalar field dominated solution allows for an inflationary evolution which is easier to attain with the sum of exponentials potential. The effective coupling of the scaling solution has an explicit dependence on the value of the coupling $C_{i\alpha}$ and the potential parameters, $\lambda_i$, for the sum of exponentials potential. For the exponential of a sum potential, however, the effective coupling can be written solely in terms of the individual couplings. We found a relation between the value of the couplings in order to obtain an early dust like dominated behaviour. Essentially, the fields must settle at the bottom of the effective potential which has a flat direction. Finally, we observed that the equations of motion for the matter density contrasts possess a source or damping  term which might lead to an unacceptable growth or damping of the density contrast. 
It would be interesting to carry out a numerical analysis to test the range of parameter space for which these models are compatible with current and forecasted future large scale structure data.

\section*{Acknowledgements}
N.J.N. was supported by the grants CERN/FP/123615/2011,
EXPL/FIS-AST/1608/2013 and PEst-OE/FIS/UI2751/2014.  L.A. acknowledges support from DFG through the project TRR33 "The Dark Universe". The authors thank David Mulryne for a careful reading of the manuscript.

\end{document}